\def\zem{$z_{\rm em}$}
\def\zabs{$z_{\rm abs}$}

\def\hi{H~{\sc i}}

\def\nhi{\mbox{$\sc N(\sc H~{\sc i})$}}
\def\lognhi{\mbox{$\log \sc N(\sc H~{\sc i})$}}

\def\feii{Fe~{\sc ii}}

\def\mgii{Mg~{\sc ii}}

\def\nii{N~{\sc ii}}

\def\nii{N~{\sc ii}}

\documentstyle[rotating]{mn}

\title[SINFONI DLAs Dynamics]{A SINFONI Integral Field Spectroscopy Survey for Galaxy Counterparts to Damped Lyman-$\alpha$ Systems - II. Dynamical Properties of the Galaxies towards Q0302$-$223 and Q1009$-$0026\thanks{Based on observations collected during programmes ESO 80.A-0330 and 80.A-0742 at the European Southern Observatory with SINFONI on the 8.2 m YEPUN telescope operated at the Paranal Observatory, Chile.} }
\author[C\'eline P\'eroux et al.] {C\'eline P\'eroux$^1$\thanks{e-mail:celine.peroux@gmail.com}, Nicolas Bouch\'e$^{2,3}$, Varsha P. Kulkarni$^4$,
Donald G. York$^5$,
\newauthor
 \& Giovanni Vladilo$^6$\\
$^1$ Laboratoire d'Astrophysique de Marseille, OAMP, Universit\'e Aix-Marseille \& CNRS,\\
38 rue Fr\'ed\'eric Joliot Curie, 13388 Marseille cedex 13, France  \\
$^2$ Max-Planck-Institut f\"ur extraterrestrische Physik Giessenbachstrasse, 85748 Garching, Germany \\
$^3$ Department of Physics, University of California, Santa Barbara, CA 93106, USA.\\
$^4$ Dept. of Physics and Astronomy, Univ. of South Carolina, Columbia, SC 29208, USA.\\
$^5$ Dept. of Astronomy and Astrophysics, Univ. of Chicago, 5640 S. Ellis Ave, Chicago, IL 60637, USA.\\
$^6$ Osservatorio Astronomico di Trieste - INAF, Via Tiepolo 11 34143 Trieste, Italy.
}
\begin{document}

%\date{Accepted 1988 December 15. Received 1988 December 14; in original form 1988 October 11}

\pagerange{\pageref{firstpage}--\pageref{lastpage}} \pubyear{2002}

\maketitle

\label{firstpage}

\begin{abstract}
Details of processes through which galaxies convert their gas into stars need to be studied in order to obtain a complete picture of galaxy formation. One way to tackle these phenomena is to relate the \hi\ gas and the stars in galaxies. Here, we present dynamical properties of Damped and sub-Damped Lyman-$\alpha$ Systems identified in H-$\alpha$ emission with VLT/SINFONI at near infra-red wavelengths. While the DLA towards Q0302$-$223 is found to be dispersion-dominated, the sub-DLA towards Q1009$-$0026 shows clear signatures of rotation. We use a proxy to circular velocity to estimate the mass of the halo in which the sub-DLA resides and find M$_{halo}$=10$^{12.6}$ M$_{\odot}$. We also derive dynamical masses of these objects, and find M$_{dyn}$=10$^{10.3}$ M$_{\odot}$ and 10$^{10.9}$ M$_{\odot}$. For one of the two systems (towards Q0302$-$223), we are able to derive a stellar mass of M$_{*}$=10$^{9.5}$ M$_{\odot}$ from Spectral Energy Distribution fit. The gas fraction in this object is 1/3$^{rd}$, comparable to similar objects at these redshifts. Our work illustrates that detailed studies of quasar absorbers can offer entirely new insights into our knowledge of the interaction between stars and the interstellar gas in galaxies.

\end{abstract}

\begin{keywords}
Galaxies:  -- galaxies: kinematics and dynamics -- quasars: absorption
   lines -- quasars: individual: Q0302$-$223, Q1009$-$0026
\end{keywords}

\section{Introduction}
%%%%%%%%%%%%%%%%%%%%%%%%%%%%%%%%%%%%%%%%%%%%%
%%%%%%%%%%%%%%%%%%%%%%%%%%%%%%%%%%%%%%%%%%%%%

Tremendous progress has been made over the last decade in establishing a broad cosmological framework in which galaxies and large-scale structure develop hierarchically over time, as a result of gravitational instabilities in the density field. The next challenge is to understand the physical processes of the formation of galaxies and structures and their interactions with the medium surrounding them. Of particular importance are the processes through which these galaxies accrete gas and subsequently form stars (Putman et al. 2009). The accretion of baryonic gas is complex. Recently, several teams (Birnboim \& Dekel 2003, Keres et al. 2005) have realized that, in halos with mass $<10^{11.5-12}$ M$_{\odot}$, baryonic accretion may not involve the traditional shock heating process of White \& Rees (1978). Similarly, details of processes through which galaxies convert their gas into stars are still poorly understood. But the observational evidences for accretion are scarce. A related signature is that the total amount of neutral gas in the Universe, $\Omega_{HI}$, is almost constant over most of the cosmic time (Prochaska, Herbert-Fort \& Wolfe 2005, P\'eroux et al. 2005, Noterdaeme et al. 2009), unlike the history of the star formation rate which peaks around z=1 (Hopkins \& Beacom 2006 and references therein). This shows the importance of ongoing global gas accretion and the conversion of atomic gas to molecular gas in the star formation process (Hopkins, Rao \& Turnshek 2005, Bauermeister et al. 2010).

One way to tackle these  problems is to relate the \hi\ gas and the stars in galaxies. While radio observations now provide detailed constraints on the \hi\ content of large sample of galaxies (Zwaan et al. 2005), they are still limited to redshift z$\sim$0. Conversely, the study of quasar absorbers, the galaxies probed by the absorption they produce in a background quasar spectrum, is insensitive to the redshift of the object (Wolfe et al. 1995). Indeed, the \hi\ content of the strongest of these quasar absorbers, the so-called Damped Lyman-$\alpha$ systems (DLAs), have been measured in samples of several hundreds of objects from the Sloan Digital Sky Survey (SDSS). However, studying the stellar content of these systems has proven very challenging until now.

In paper I of these series (P\'eroux et al. 2010), we reported the detection at infra-red wavelengths of H-$\alpha$ emission of a Damped Lyman-$\alpha$ (DLA) system at \zabs=1.009 and one sub-Damped Lyman-$\alpha$ system at \zabs=0.887 with \nhi$>$10$^{19}$ (P\'eroux et al. 2003). Here, we take advantage of 3D spectroscopy made possible by the SINFONI instrument on VLT to study the spatially resolved kinematics and therefore the dynamical state of these galaxies. In all the remaining, we assumed a cosmology with H$_0$=71 km/s/Mpc, $\Omega_M$=0.27 and $\Omega_{\rm \lambda}$=0.73.

\section{SINFONI Detections of H-$\alpha$ Emission}
\begin{table*}
\begin{center}
\caption{Summary of the properties of two absorber galaxies detected with SINFONI. }
\label{t:results}
\begin{tabular}{ccccccccc}
\hline\hline
Quasar 		  &\zabs &\lognhi &[Zn/H] &impact parameter &$\Delta v^a$ &F(H-$\alpha$) &Lum(H-$\alpha$) &SFR $^b$  \\
 		    &&[atoms/cm$^2]$ &&[kpc] &[km/s]&[erg/s/cm$^2]$ &[erg/s]&[M$_{\odot}$/yr]  \\
\hline
Q0302$-$223	     &1.009 &20.36$^{+0.11}_{-0.11}$  &$-$0.51$\pm$0.12    &25 &1 &7.7$\pm$2.7$\times$10$^{-17}$&4.1$\pm$1.4$\times$10$^{41}$&1.8$\pm$0.6 \\
Q1009$-$0026	     &0.887	&19.48  $^{+0.05}_{-0.06}$ &$+$0.25$\pm$0.06  &39  &125 &17.1$\pm$6.0$\times$10$^{-17}$ &6.6$\pm$2.3$\times$10$^{41}$&2.9$\pm$1.0\\
\hline\hline 				       			 	 
\end{tabular}			       			 	 
\end{center}			       			 	 
\vspace{0.2cm}
\begin{minipage}{140mm}
{\bf $^a$:} Observed velocity shift between the \zabs\ and \zem\ of the detected galaxies. \\
{\bf $^b$:} The SFR estimates are not corrected for dust extinction.\\
\end{minipage}
\end{table*}			       			 	 

The observations presented here were carried out at the European Southern Observatory with the near-infrared field spectrometer SINFONI on Unit 4 of the Very Large Telescope. First results with that instrument reported a high success rate in detecting metal absorption systems traced by \mgii\ absorbers (Bouch\'e et al 2007a) at $\sim$1, while lower detection rates are found at higher-redshifts z$\sim$2 (Bouch\'e et al., private communication). In Paper I (P\'eroux et al. 2010), we have reported two secure detections of the redshifted H-$\alpha$ emission of high \hi\ column density quasar absorbers: a DLA with log \nhi=20.36$\pm$0.11 at \zabs=1.009 towards Q0302$-$223 and a sub-DLA with log \nhi=19.48$^{+0.05}_{-0.06}$ at \zabs=0.887 towards Q1009$-$0026 (see Table~\ref{t:results}). We detect galaxies associated with the quasar absorbers at impact parameters of 25 and 39 kpc away from the quasar sightlines, respectively. 

For field Q0302$-$223 where the quasar is bright enough, we have used the quasar itself as a natural guide star for adaptive optics in order to improve the spatial resolution (see Table~\ref{t:results}). The two data sets have a resulting Point Spread Function (PSF) of 0.7" and 1.0". Both the objects in our study have well-determined absorption line properties determined from spectra of the background quasars. Thus, the metallicities are well-determined for both of our target absorbers. An estimate of the emission metallicities is made based on the \nii / H-$\alpha$ parameter (Pettini \& Pagel 2004): the galaxy with the smaller absorption line metallicity also has the smaller emission-line metallicity. The absorber toward Q0302$-$223 is the more gas-rich absorber, while that toward Q1009$-$0026 is the more metal-rich absorber.
Using the H-$\alpha$ luminosity we then derived the star formation rate assuming the Kennicutt (1998) flux conversion corrected to a Chabrier (2003) initial mass function. We find low star formation rates (not corrected for dust extinction), of 1.8$\pm$0.6 and 2.9$\pm$1.0 M$_{\odot}$/yr (see Table~\ref{t:results}). These values of star formation rates are among the lowest that have ever been possible to detect in quasar absorber searches with ground-based observations at z$\sim$1 and 2.

\section{Kinematics}

\begin{figure*}
\begin{center}
\includegraphics[height=5.5cm, width=7cm, angle=-90]{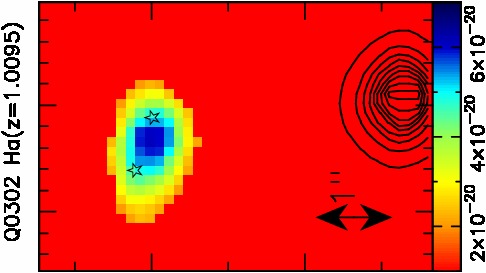}
\includegraphics[height=5.5cm, width=7cm, angle=-90]{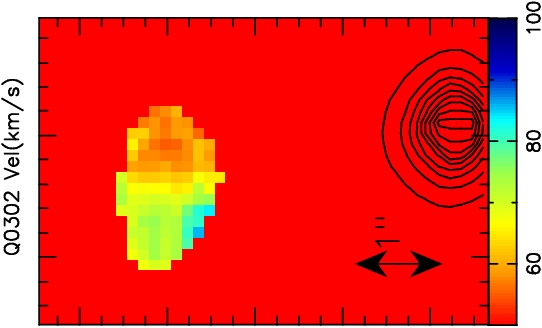}
\includegraphics[height=5.5cm, width=7cm, angle=-90]{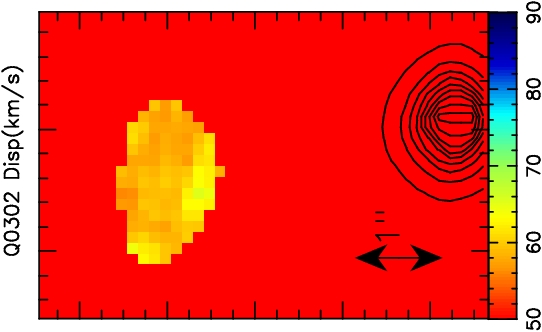}
\includegraphics[height=5.75cm, width=6cm, angle=-90]{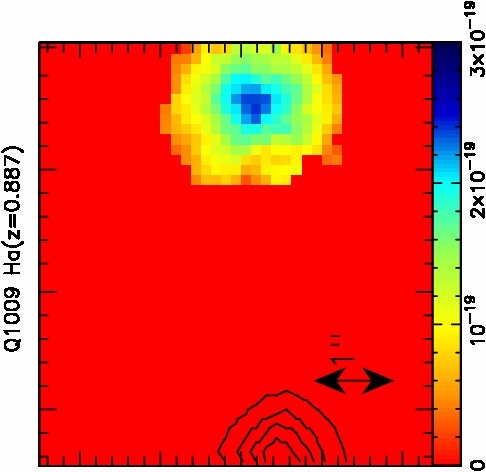}
\includegraphics[height=5.75cm, width=6cm, angle=-90]{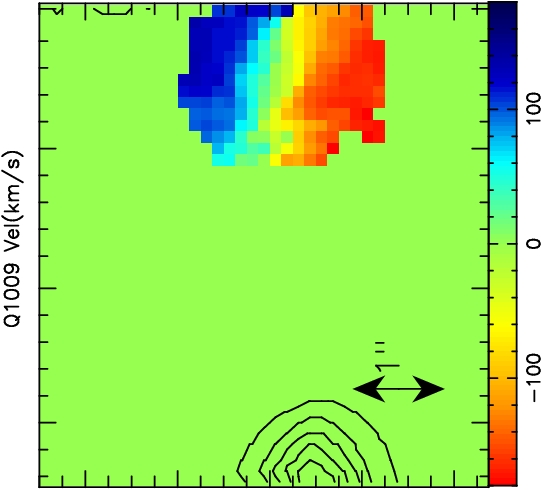}
\includegraphics[height=5.75cm, width=6cm, angle=-90]{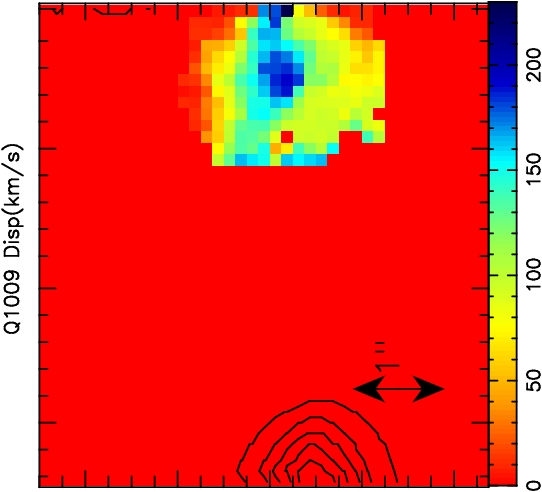}
\caption{{\bf H-$\alpha$ flux map, H-$\alpha$ velocity field and H-$\alpha$ velocity dispersion maps of the two quasar absorbers.} The colour-scale indicates the flux in erg/s/cm$^2$ in the H-$\alpha$ flux map on the left and the velocities or velocity dispersions in km/s for the middle and right set of panels. North is up and east is to the left. The thin lined, black contours indicate the position of the quasar and the arrow represents 1 arcsec which corresponds to 8.1 kpc in the case of Q0302$-$223 and 7.8 kpc in the case of Q1009$-$0026. The system at the top (Q0302$-$223) is dispersion-dominated with H-$\alpha$ luminosity consistent with that of a disk. The system at the bottom (Q1009$-$0026) shows a blueshifted and redshifted component of the gas on its velocity map. This pattern is a clear characteristic of a rotating disk.}
\label{f:vel}
\end{center}
\end{figure*}

In addition to the identification and redshift confirmation of the galaxy responsible for the quasar absorbers, the SINFONI data allow for a study of the dynamical properties of the galaxies. We extracted maps of the velocity-integrated line fluxes, relative velocities, and velocity dispersion from the reduced data cubes using the following technique. For each pixel, we fitted the line profile to convolved gaussian with a template of the instrument profile following e.g., F\"orster-Schreiber et al. (2006, 2009). Thus, the fitting takes into account the intrinsic emission line profile. In addition, weighted fits are performed using the MC resampling based on the spectral noise profile. A boxcar median smooth of size 2$\times$2 pixels was applied prior to the line fitting. The velocity resolution is 150 km/s while the resulting accuracy of the centroid of the line can be derived to 0.1 of a pixel (29 km/s). These maps are displayed in Figure~\ref{f:vel}.

Thanks to these, critical information on the dynamical state and properties of galaxies associated with quasar absorbers can be derived. The galaxy associated with the \zabs=1.009 DLA towards Q0302$-$223 shows little signs of rotation and significant amounts of dispersions. It has an axis ratio corrected for Point Spread Function (PSF) of b/a=0.47 corresponding to an inclination of sin $i$=0.88. The maximum velocity extracted from the data cube is 2 V$_{max}$ sin i= 20 km/s, corresponding to an inclination corrected maximum velocity of V$_{max}$=11 km/s. Given that the velocity dispersion (corrected from instrumental effects) is 59 km/s, the ratio v/$\sigma$ in this object is 0.19 (see Table~\ref{t:kine}), much smaller than seen in local disk galaxies (v/$\sigma$=10-20), and at high-redshifts (Epinat et al. 2009, Cresci et al. 2009, Lemoine-Busserolle et al. 2010). This might indicate that this object is pressure-supported, and early-type morphologies come to mind.

However, the morphology of this object indicates that this galaxy might be a late type object. The SINFONI data shows that it is highly elongated. This is supported by the archival HST/WFPC2 images covering this field (Le Brun et al. 1997). In these data, two separate components can be seen but their redshifts cannot be estimated (see paper I and the two stars in Figure~\ref{f:vel} which indicate the positions of the components). The two individual components are not resolved in the SINFONI observations {and might explain the elongated shape or only one is seen but the detection of H-$\alpha$ emission at the position of the absorber provides secure identification. On interpretation is that we might be seeing a dispersion due to a small difference in redshift of two interacting galaxies. However, the H-$\alpha$ light profile appears to be exponential with a scale length of 0.6". Thus, alternative explanation is that this galaxy is dispersion dominated, albeit it already has a disky morphology. This is indicative of a young flattened object, with effective radius (containing half of the light) of R$_{e}$=0.5" or 4 kpc, whose kinematics show that sustained rotation has not yet occured. In other words, this object will have to contract further. 

By contrast, the absorber toward Q1009$-$0026 has a morphology and kinematics consistent with that of a disk, with a normal dispersion profile
(v/$\sigma$=1.45) peaking at the center $\sigma_{peak}$=190 km/s and flattening out in the outer-parts $\sigma_{disk}$=60-70 km/s (see Table~\ref{t:kine}). This object is more face-on with an estimated inclination of sin $i$=0.60 derived from an axes ratio b/a=0.80. It shows clear signatures of rotation with systematic velocity gradients.  Its v/$\sigma$ is not typical of local disk galaxies which have v/$\sigma$=10-20, but the systematic gradient still favors a spiral galaxy. 

We also use the kinematic maps to estimate the size of the systems, correcting for the seeing in quadrature. These, in turn, are used to compute the area of the objects assuming that the inclined disks appear as ellipses. Using these estimates of the sizes of the detected galaxies, we compute their star formation rate surface densities (see Table~\ref{t:kine}) and find $\Sigma_{SFR}$=0.13 and 0.31 solar masses per year per kpc squared for absorbers towards Q0302$-$223 and Q1009$-$0026 respectively. It is interesting to note that these values are above the threshold at 0.1 solar masses per kpc squared where galactic winds are observed (Heckman 2002).  In the log $\Sigma_{gas}$-log $\Sigma_{SFR}$ plot of Bouch\'e et al. (2007b), our galaxies appear to be similar to the z$\sim$1.5 colour-selected galaxies from the Gemini Deep Deep Survey (Abraham et al. 2004), i.e. at the lower end of the relation with respect to higher-redshift galaxies and submillimeter-selected systems. The results of these are shown in Table~\ref{t:kine}.

\section{Mass Estimates}

\begin{table*}
\begin{center}
\caption{Kinematic properties and mass estimates of the two \nhi\ absorbers detected (see text for details of the computations). }
\label{t:kine}
\begin{tabular}{cccccccccc}
\hline\hline
Quasar 		  &sin $i$ &v/$\sigma$   &r$_{1/2}$ & $\Sigma_{SFR}$  &$M_{dyn}$      &$\Sigma_{gas}$	&$M_{gas}$	& $M_{halo}$& $M_{*}$ \\
		           &          &	&["]&[M$_{\odot}$/yr/kpc$^2$] &[M$_{\odot}$] 	&[M$_{\odot}$/pc$^2$]       &[M$_{\odot}$]        &[M$_{\odot}$] &[M$_{\odot}$]  \\
\hline
Q0302$-$223{\bf $^a$}	&0.88	&0.19	&0.7 &0.13	&10$^{10.3}$	&10$^{1.9}$	&10$^{9.1}$	&--	&10$^{9.5}$	  \\
Q1009$-$0026	     		&0.60	&1.45	&0.5 &0.31	&10$^{10.9}$	&10$^{2.2}$	&10$^{9.2}$	&10$^{12.6}$&--    \\
\hline\hline 				       			 	 
\end{tabular}			       			 	 
\end{center}			       			 	 
\vspace{0.2cm}
\begin{minipage}{140mm}
{\bf Note:} The inclination is the main source of uncertainties and is estimated to be around 30\%.\\
{\bf $^a$:} The higher-resolution HST/WFPC2 data from Le Brun et al. (1997) clearly shows that the object is subdivided into two sub-components, consistent with the elongated shape seen in the SINFONI data presented here. In this table, however, the object is treated as only one. \\
\end{minipage}
\end{table*}

{\it Dynamical Masses:} These observations provide direct observational estimates of dynamical masses for galaxies selected on neutral gas HI content. For Q0302$-$223, we use the virial theorem to calculate the total mass of the system since the rotation is not established yet (Wright et al. 2009):

\begin{equation}
M_{\rm dyn} = C \sigma^2~r_{1/2}~/~G
\end{equation}

where C is a constant (C=5 for a spherically uniform density profile and r$_{1/2}$ is the half-light radius. We derive r$_{1/2}$ from a curve of growth with ellipsoidal apertures. 
We find that r$_{1/2}$=0.7" (corrected for seeing) and hence M$_{dyn}$=10$^{10.3}$ solar masses for Q0302$-$223.

In the case of Q1009$-$0026, we know that the system is rotating and we can use the enclosed mass to determine the dynamical mass (Epinat et al. 2009):

\begin{equation}
M_{\rm dyn} = V_{max}^2~r_{1/2}~/~G
\end{equation}

where $V_{max}$ is the maximum velocity and r$_{1/2}$ is as before. Estimating  r$_{1/2}$ from a curve of growth is more difficult in this case because of the proximity of the galaxy to the edge of SINFONI's field of view. As a consequence, we measured R$_{hwhm }$, the half width at halfmax, from a Gaussian fit to the 1D profile extracted along the major kinematic axis.  We find it to be R$_{hwhm }$=0.64", corresponding to r$_{1/2}$ of $\sim$0.8" using the empirical relation r$_{1/2}$=1.25 x R$_{hwhm }$ for disks and under a seeing of 1" (Bouch\'e et al. 2007b; Bouch\'e in preparation). We therefore find M$_{dyn}$=10$^{10.9}$ solar masses for Q1009-0026 (see Table~\ref{t:kine}).

{\it Mass of Gas:} In order to estimate the mass of gas in these objects, we start from the observed H-$\alpha$ surface brightness and compute gas surface brightness using an inverse "Schmidt-Kennicutt" relation (Bouch\'e et al. 2007b, Finkelstein et al. 2009):

\begin{equation}
\Sigma_{gas} [M_{\odot}/pc^{2}] = 1.6 \times 10^{-27}\left(\frac{\Sigma_{H\alpha}}{[erg/s/kpc^{2}]}\right)^{0.71} 
\end{equation}

Gas masses inferred in this way are subject
to significant uncertainties beyond those associated with the measurements
of fluxes and sizes and the 0.3 dex uncertainty introduced
by the scatter in the Schmidt-Kennicutt law itself. It is not yet
known whether the Schmidt-Kennicutt law in this form holds at high redshift. 
 In addition, the calculation is based on integrating 
the gas surface density over the visible 
region, but more gas could lie beyond this limit as in the case of the Milky Way.
In the galaxy in the field of quasar Q0302$-$223, we find a gas mass M$_{gas}$=10$^{9.1}$ solar masses and we find 10$^{9.2}$ in the other system. In comparison with the dynamical masses M$_{dyn}$ above, these results indicate a low gas fraction in the objects. In addition, both objects show clear exponential light profiles indicative of disks.

{\it Halo Mass:} Thanks to our kinematical data based on the H-$\alpha$ emission widths (reflecting the velocity dispersion of the warm gas), we are able to estimate the mass of the halo in which the system towards Q1009$-$0026 resides, assuming a spherical virialised collapse model (Mo \& White 2002):

\begin{equation}
M_{\rm halo}= 0.1 H_o^{-1} G^{-1} \Omega_m^{-0.5} (1+z)^{-1.5} V_{max}^3
\end{equation}

using inclination-corrected V$_{max}$ values. We find M$_{halo}$=10$^{12.6}$ solar masses. This halo mass is comparable with the one from the Milky Way: 1.9$^{+3.6}_{-1.7} \times$ 10$^{12}$ M$_{\odot}$ (Wilkinson et al. 1999).

\section{Spectral Energy Distribution Fit}

\begin{figure}
\begin{center}
\includegraphics[height=9cm, width=6cm, angle=-90]{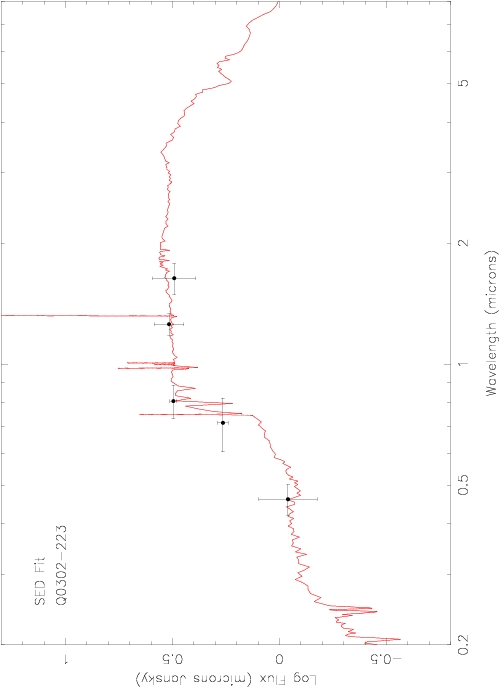}
\caption{SED fit of emission line galaxy to the broad-band magnitudes from HST/WFPC2 (Le Brun et al. 1997) and ground-based near-infrared observations (Chen \& Lanzetta 2003) based on "Le Phare" photometric-redshift code (Ilbert et al. 2009). The spectroscopic redshift and star formation rate derived from our SINFONI spectra is used as an input to the code, thus allowing one to constrain the stellar mass of the object with relative high confidence, M$_{*}$=10$^{9.5}$ M$_{\odot}$ and the age, about 2.5 Gyr. }
\label{f:SED_spectrum}
\end{center}
\end{figure}

{\it Stellar Mass:} In the case of the DLA towards Q0302$-$223, we use the broad-band magnitudes from HST/WFPC2 (Le Brun et al. 1997) and ground-based near-infrared observations (Chen \& Lanzetta 2003) which cover the Balmer break of the object to constrain the age of the stellar population in the galaxy with a Spectral Energy Distribution (SED) fit to the integrated light of the galaxies. We used the photometric-redshift code "Le Phare" (Ilbert et al. 2009), with a single burst of star formation, templates from Bruzual \& Charlot (2003) stellar population models, a Calzetti extinction law and a Charbrier (2003) initial mass function. The spectroscopic redshift and star formation rate derived from our SINFONI spectra are used as an input to the code, thus allowing one to constrain the stellar mass of the object with relatively high confidence. We deduce M$_{*}$=10$^{9.5}$ M$_{\odot}$ for an object that would be about 2.5 Gyr old. 
An illustration of the template which best fit the observed magnitudes is provided in Figure~\ref{f:SED_spectrum}.

This, in turn, allow us to put constraints on the baryonic mass fraction in this galaxy:

\begin{equation}
f_{baryons}=(M_{gas}+M_{*})/M_{dyn}
\end{equation}
 
We derive a baryonic fraction of the order of 20\%. Similarly, we can estimate the gas fraction is this object:

\begin{equation}
\mu=M_{gas}/(M_{gas}+M_{*})
\end{equation}

We derive a gas fraction of 1/3$^{rd}$. Such gas fractions are in the low range of the typical values derived in z$\sim$2-3 galaxies by others (e.g. Erb et al. 2006, Law et al. 2009). When comparing these various mass estimates, we see that these systems have 
little room for molecular gas, which is consistent with the low star formation rates derived.

\section{Comparison with Absorption Properties}

What makes our targets unique in comparison with other galaxies at z$\sim$1 observed with Integral Field Unit (e.g. F\" orster-Schreiber et al. 2009, Law et al. 2009) is the wealth of information about the absorption properties of the gas available from high-resolution Echelle spectroscopy (see also Kacprzak et al. 2010). This allows us to compare the distribution of velocities of the neutral gas (absorption) and the ionised gas (emission). In the case of the absorber towards Q0302$-$223, Pettini et al. (2000) have fit the absorption profile with two main groups of components, separated by $\Delta v =$36 km/s. They also indicate additional weaker components, at v=35 and 121 km/s relative to \zabs=1.009, which are visible in the stronger \feii\ lines. 
n the case of the sub-DLA towards Q1009$-$0026, Meiring et al. (2006) have reported  complicated velocity structure with seven components spanning $\Delta v \sim$334 km/s. This is considerably larger than the velocity dispersion measured at the position where it flattens out in the velocity map, $\sigma_{disk}$=60-70 km/s. This may support previous claims that the width of the absorption systems cannot be solely explained by gravitational motions but by the presence of outflows (Bouch\'e et al. 2007a). 

\section{Conclusion}

In conclusion, the observational set-up of SINFONI has demonstrated the power of integral field spectroscopy for deriving a number of emission properties for quasar absorbers, a type of high-redshift galaxies that have been difficult to identify in the past. Detailed dynamical properties of these galaxies with known gas characteristics could be derived. These new tools are now available to systematically study these objects. We find that the two absorbers studied here have kinematical properties similar to other galaxies studied in the same way at these and others redshifts (Epinat et al. 2009, F\" orster-Schreiber et al. 2006, Genzel et al. 2006, F\" orster-Schreiber et al. 2009). 

We find that galaxy associated with the DLA identified towards Q0302$-$223 shows little signs of rotation and significant amounts of dispersions. Moreover, the light profile and morphology  provide further evidence that this galaxy is dispersion dominated, albeit it already has a disky morphology. This is indicative of a young object which is confirmed by results from a SED fit or could be due to the blending of the two components detected with HST interacting together.  On the contrary, the galaxy associated with the sub-DLA towards Q1009$-$0026 has a morphology and kinematics consistent with that of a disk, with a normal dispersion profile peaking at the center and flattening out in the outer-parts. This object shows clear signatures of rotation with systematic velocity gradients which is not typical of local disk galaxies, but the systematic gradient still favors a spiral galaxy. It remains to be understood why such high \nhi\ column densities are found at large impact parameters (up to b=39kpc) from galaxies.

Overall, we conclude that of the two absorbers studied, the less metal-rich absorber toward Q0302$-$223 arises in a gas-rich system with H-$\alpha$ luminosity consistent with that of a disk, while the more metal-rich absorber toward Q1009$-$0026 arises in a lower gas mass system but with higher total mass showing clear signs of rotation. With the limitations of small number statistics, our findings are consistent with the suggestion (Khare et al. 2007; Kulkarni et al. 2007) that the metal-rich sub-DLAs may arise in more massive galaxies compared to the DLAs, which are usually metal-poor by comparison. Our work illustrates that detailed studies of quasar absorbers can offer entirely new insights into our knowledge of the interaction between stars and the interstellar gas in galaxies.

\section*{Acknowledgements}
We would like to thank the Paranal and Garching staff at ESO for performing the observations in Service Mode and Olivier Ilbert for performing the SED fit. 
VPK acknowledges partial support from the U.S. National Science Foundation grants AST-0607739 and AST-0908890 (PI: Kulkarni). This work has benefited from support of the "Agence Nationale de la Recherche" with reference ANR-08-BLAN-0316-01.

\bsp

\label{lastpage}


\begin{thebibliography}{99}


 \bibitem[]{} Abraham, R. G., Glazebrook, K., McCarthy, P. J., Crampton, D., Murowinski, R., Jorgensen, I., Roth, K., Hook, I. M., Savaglio, S., Chen, H-W., Marzke, R. O. \& Carlberg, R. G., 2004, AJ, 127, 2455
 
 \bibitem[]{} Bauermeister, A., Blitz, L. \& Ma, C., 2010, ApJ, 717, 323
 
 \bibitem[]{} Birnboim, Y. \& Dekel, A., 2003, MNRAS, 345, 349

 \bibitem[]{} Bouch\'e, N., Murphy, M. T.,  P\'eroux, C., Davies, R., Eisenhauer, F., F\"orster Schreiber, N. M. \& Tacconi, L., 2007a, ApJ, 669L, 5 

\bibitem[]{} Bouch\'e, N., Cresci, G., Davies, R., Eisenhauer, F., F\" orster Schreiber, N. M., Genzel, R., Gillessen, S., Lehnert, M., Lutz, D., Nesvadba, N., Shapiro, K. L., Sternberg, A., Tacconi, L. J., Verma, A., Cimatti, A., Daddi, E., Renzini, A., Erb, D. K., Shapley, A., Steidel, C. C. 2007b, ApJ, 671, 303

\bibitem[]{} Bruzual, G. \& Charlot, S., 2003, MNRAS, 344, 1000

\bibitem[]{} Charbrier, C.,  2003, ApJ, 586L, 133

\bibitem[]{} Chen, H.-W. \& Lanzetta, K. M., 2003, ApJ, 597, 706

\bibitem[]{} Cresci, G., et al., 2009, ApJ, 697, 115

\bibitem[]{} Epinat, B., Contini, T., Le Fevre, O., Vergani, D., Garilli, B., Amram, P., Queyrel, J., Tasca, L. \& Tresse, L., 2009, A\&A, 504, 789

\bibitem[]{} Erb, D. K., Steidel, C. C., Shapley, A. E., Pettini, M., Reddy, N. A. \& Adelberger, K. L., 2006, ApJ, 646, 107

\bibitem[]{} Finkelstein, S. L., Papovich, C., Rudnick, G., Egami, E., LeFloc'h, E., Rieke, M. J., Rigby, J. R. \&  Willmer, C. N. A., 2009, ApJ, 700, 376

\bibitem[]{} F\"orster Schreiber, N. M., Genzel, R., Lehnert, M. D., Bouch\'e, N., Verma, A., Erb, D. K., Shapley, A. E., Steidel, C. C., Davies, R., Lutz, D., Nesvadba, N., Tacconi, L. J., Eisenhauer, F., Abuter, R., Gilbert, A., Gillessen, S. \& Sternberg, A., 2006, ApJ, 645, 1062

\bibitem[]{} F\"orster Schreiber, N. M., Genzel, R., Bouch\'e, N., Cresci, G., Davies, R., Buschkamp, P., Shapiro, K., Tacconi, L. J., Hicks, E. K. S., Genel, S., Shapley, A. E., Erb, D. K., Steidel, C. C., Lutz, D., Eisenhauer, F., Gillessen, S., Sternberg, A., Renzini, A., Cimatti, A., Daddi, E., Kurk, J., Lilly, S., Kong, X., Lehnert, M. D., Nesvadba, N., Verma, A., McCracken, H., Arimoto, N., Mignoli, M., Onodera, M., 2009, ApJ, 706, 1375

\bibitem[]{} Heckman, T., 2002, Extragalactic Gas at Low Redshift, 254, 292

\bibitem[]{} Hopkins, A. M., Rao, S. M. \& Turnshek, D. A., 2005, ApJ, 630, 108

\bibitem[]{} Hopkins, A. M. \& Beacom, J. F.,  2006, ApJ, 651, 142

\bibitem[]{} Ilbert, O. et al., 2009, ApJ, 690, 1236

\bibitem[]{} Kacprzak, G. G., Churchill, C. W., Ceverino, D., Steidel, C. C., Klypin, A. \& Murphy, M. T.,  2010, ApJ, 711, 533

\bibitem[]{} Kennicutt, R. C., Jr., 1998, ApJ, 498, 541

\bibitem[]{} Keres, D., Katz, N.,  Weinberg, D. H. \& Dav\'e, R., 2005, MNRAS, 363, 2

\bibitem[]{} Khare, P., Kulkarni, V. P., P\'eroux, C., York, D. G., Lauroesch, J. T. \& Meiring, J. D., 2007, A\&A, 464, 487

\bibitem[]{} Kulkarni, V. P., Khare, P., P\'eroux, C., York, D. G., Lauroesch, J. T. \& Meiring, J. D., 2007, ApJ, 661, 88

\bibitem[]{} Law, D. R., Steidel, C. C., Erb, D. K., Larkin, J. E., Pettini, M., Shapley, A. E. \& Wright, S. A., 2009, ApJ, 697, 2057 

\bibitem[]{} Le Brun, V., Bergeron, J., Boisse, P., Deharveng, J. M., 1997, A\&A, 321, 733

\bibitem[]{} Lemoine-Busserolle, M., Bunker, A., Lamareille, F. \& Kissler-Patig, M., 2010, MNRAS, 401, 1657

\bibitem[]{} Meiring, J. D., Kulkarni, V. P., Khare, P., Bechtold, J., York, D. G., Cui, J., Lauroesch, J. T., Crotts, A. P. S. \& Nakamura, O., 2006, MNRAS, 370, 43
 
\bibitem[]{} Mo, H. J. \& White, S. D. M.,  2002, MNRAS, 336, 112 

\bibitem[]{} Noterdaeme, P., Petitjean, P., Ledoux, C., Srianand, R., 2009, A\&A, 505, 1087

\bibitem[]{} P\'eroux, C., McMahon, R. G., Storrie-Lombardi, L. J. \& Irwin, M., 2003, MNRAS, 346, 1103

\bibitem[]{} P\'eroux, C., Dessauges-Zavadsky, M., D'Odorico, S., Kim, T. S. \& McMahon, R. G., 2005, MNRAS, 363, 479

\bibitem[]{} P\'eroux, C., Bouch\'e, N., Kulkarni, V. P., York, D. G. \& Vladilo, G.  2010, submitted (Paper I)

\bibitem[]{} Pettini, M., Ellison, S. L., Steidel, C. C., Shapley, A. E. \& Bowen, D. V., 2000, ApJ, 532, 65

\bibitem[]{} Pettini, M \& Pagel, B. E. J., 2004, MNRAS, 348L, 59

\bibitem[]{} Prochaska, J. X., Herbert-Fort, S. \& Wolfe, A. M., 2005, ApJ, 635, 123

\bibitem[Putman et al. 2009]{Putman}Putman, M.E., et al. 2009, "How do galaxies accrete gas and form stars?", The Astronomy \& Astrophysics Decadal Survey 2010 

\bibitem[]{} White, S. D. M. \& Rees, M. J., 1978, MNRAS, 183, 341

\bibitem[]{} Wilkinson, M. I. \& Evans, N. W., 1999, MNRAS, 310, 645

\bibitem[]{} Wolfe, A. M., Lanzetta, K. M., Foltz, C. B. \& Chaffee, F. H., 1995, ApJ, 454, 698

\bibitem[]{} Wright, S. A., Larkin, J. E., Law, D. R.; Steidel, C. C.; Shapley, A. E. \& Erb, D. K., 2009, ApJ, 699, 421 

\bibitem[]{} Zwaan, M. A., Meyer, M. J., Staveley-Smith, L.; \& Webster, R. L., 2005, MNRAS, 359L, 30

\end{thebibliography}
\end{document}